\documentclass[11pt]{article}
\usepackage{amsmath}
\usepackage{amsfonts}
\usepackage{amssymb}
\usepackage{graphicx}
\usepackage{bm}
\usepackage{color}
\usepackage{amscd}

\def\bea{\begin{eqnarray}}
\def\eea{\end{eqnarray}}

\begin{document}
\begin{center}
\LARGE {\bf Horizon Fluffs: In the Context of Generalized Minimal Massive Gravity}
\end{center}

\begin{center}
{M. R. Setare \footnote{E-mail: rezakord@ipm.ir}\hspace{1mm} ,
H. Adami \footnote{E-mail: hamed.adami@yahoo.com}\hspace{1.5mm} \\
{\small {\em  Department of Science, University of Kurdistan, Sanandaj, Iran.}}}\\

\end{center}

\begin{center}
{\bf{Abstract}}\\
We consider a metric which describes Ba$\tilde{\text{n}}$ados geometries and show that the considered metric is a solution of generalized minimal massive gravity (GMMG) model. We consider the Killing vector field which preserves the form of considered metric. Using the off-shell quasi-local approach we obtain the asymptotic conserved charges of given solution. Similar to the Einstein gravity in the presence of negative cosmological constant, for the GMMG model also, we show that the algebra among the asymptotic conserved charges is isomorphic to two copies of the Virasoro algebra. Eventually, we find relation between the algebra of the near horizon and the asymptotic conserved charges. This relation show that the main part of the horizon fluffs proposal of Refs.\cite{140,14} appear for generic black holes in the class of Ba$\tilde{\text{n}}$ados geometries in the context of GMMG model.
\end{center}

\section{Introduction}\label{A}
The near horizon symmetries of the black hole solutions in three dimensions are related with the Bondi–-van der Burg–-Metzner–-Sachs (BMS) algebra \cite{5'}. The BMS symmetry algebra in $n$ space-time dimension consists
 of the semi-direct sum of the conformal Killing vectors of a $(n-2)$-dimension sphere acting on the ideal of infinitesimal
 supertranslations \cite{4',5''}. Recently, Donnay et al \cite{11'}, have shown that the asymptotic symmetries close to the horizon of the non-extremal black hole solution of the three-dimensional Einstein gravity in the presence of a negative cosmological term, are generated by an extension of supertranslations. They have shown that for a special choice of boundary conditions, the near region to the horizon of a stationary black hole presents a generalization of supertranslation, including a
semidirect sum with superrotations, represented by Virasoro algebra. More recently, we have studied the behaviors and algebras of the symmetries and conserved charges near the horizon of the non-extremal black holes in the context of the so-called Generalized Minimal Massive Gravity \cite{8}, proposed in Ref. \cite{7}. The authors of \cite{19} have studied the spacetime geometry around the non-extermal horizon in the context of the Einstein gravity in the presence of negative cosmological constant. They have proposed a new set of the boundary conditions which leads to a near horizon symmetry, the Heisenberg algebra. In another paper we have studied this near horizon symmetry in the framework of Chern-Simons-like theories of gravity \cite{3}. In another term, similar to the near horizon symmetry algebra of the black flower solutions of the Einstein gravity in the presence of negative cosmological constant, we have found an algebra consists
of two $U(1)$ current algebras, but instead with levels $\pm\frac{k}{2}$, the level of our algebra is given by $\mp \frac{k}{2} \left( \sigma \pm \frac{1}{\mu l}+ \frac{\alpha H}{\mu} + \frac{F}{m^{2}} \right)$.  As has been discussed in Refs. \cite{140,14}, in field theories with local symmetries, there are conserved charges due to the residual gauge symmetries, which makes the field configuration distinguishable. In \cite{140} the authors have found a microcanonical description of the black hole microstates
as "near horizon soft hairs". Then they have studied the "horizon fluffs" proposal of \cite{140} for generic black
holes in the class of Banados geometries in \cite{14}. They have done this study in the context of the Einstein gravity in the presence f negative cosmological constant. The present our work generalize the analysis done in the case of the Einstein gravity to a class of Chern-Simons-like, higher curvature theories. We know that the pure Einstein-–Hilbert gravity in three dimensions (in the presence of negative cosmological constat also)  exhibits no propagating physical degrees of freedom \cite{2',3'}.  Adding the gravitational Chern-Simons term produces a
propagating massive graviton \cite{1}. The resulting
theory is called topologically massive gravity (TMG). Unfortunately TMG has
a bulk-boundary unitarity conflict. In another term either the bulk or the
boundary theory is non-unitary, so there is a clash between the positivity
of the two Brown-Henneaux boundary $c$ charges and the bulk energies \cite{1007}. But as has been shown in \cite{7}, fortunately GMMG  avoids the aforementioned ``bulk-boundary unitarity clash''.
Calculation of the GMMG action to quadratic order about $AdS_3$ space show that the theory is free of negative-energy bulk modes. Also Hamiltonian analysis show that the GMMG model has no Boulware-Deser ghosts and this model propagate only two physical modes.
 So this model is viable candidate for semi-classical limit of a unitary quantum $3D$ massive gravity. So this is our reason and motivation to generalize the Einstein gravity studies in \cite{140,14} to higher curvature theories in the presence of Chern-Simons term. But totally, the motivation for the above mentioned recent studies of near horizon symmetries is due to the the recent nice work of Hawking et al \cite{180}, because people think may be we can find a solution to the information paradox for black holes by analysis of near horizon symmetries. \\
The paper is organised as follows. In Sec.\ref{B}, we briefly review the method of obtaining conserved charges of CSLTG by the off-shell quasi-local approach. Then we introduce the GMMG model as an example of CSLTs of gravity. In Sec.\ref{C} we consider the Ba$\tilde{\text{n}}$ados geometries and their symmetries. In Sec.\ref{D} we obtain the conserved charges of asymptotic and near horizon regions, then present their algebras. Sec .\ref{E} is devoted to conclusions and discussions.
\section{Quasi-local conserved charges of generalized minimal massive gravity}\label{B}
\subsection{Conserved charges in the context of CSLTG}\label{B.1}
Let's start by reviewing the method of obtaining conserved charges of Chern-Simons-like theories of gravity (CSLTG) by the  off-shell quasi-local approach \cite{3}. The Lagrangian 3-form of the Chern-Simons-like theories of gravity is given by \cite{1}
\begin{equation}\label{1}
  L=\frac{1}{2} \tilde{g}_{rs} a^{r} \cdot da^{s}+\frac{1}{6} \tilde{f}_{rst} a^{r} \cdot a^{s} \times a^{t}.
\end{equation}
In the above Lagrangian $ a^{ra}=a^{ra}_{\hspace{3 mm} \mu} dx^{\mu} $ are the Lorentz vector valued one-forms, where, $r=1,...,N$ and $a$ indices refer to flavour and the Lorentz indices, respectively. We should mention that, here, the wedge products of the Lorentz-vector valued one-form fields are implicit. Also, $\tilde{g}_{rs}$ is a symmetric constant metric on the flavour space and $\tilde{f}_{rst}$ is a totally symmetric "flavour tensor" which are interpreted as the coupling constants. We use a 3D-vector algebra notation for the Lorentz vectors in which contractions with $\eta _{ab}$ and $\varepsilon ^{abc}$ are denoted by dots and crosses, respectively \footnote{Here we consider the notation used in \cite{1}.}. It is worth saying that $a^{ra}$ is a collection of the dreibein $e^{a}$, the dualized spin-connection $\omega ^{a}$, the auxiliary field $ h^{a}_{\hspace{1.5 mm} \mu} = e^{a}_{\hspace{1.5 mm} \nu} h^{\nu}_{\hspace{1.5 mm} \mu} $ and so on. Also for all interesting CSLTGs we have $\tilde{f}_{\omega rs} = \tilde{g}_{rs}$ \cite{2}.\\
The total variation of $a^{ra}$ due to a diffeomorphism generator $\xi$ is \cite{4}
\begin{equation}\label{2}
  \delta _{\xi} a^{ra} = \mathfrak{L}_{\xi} a^{ra} -\delta ^{r} _{\omega} d \chi _{\xi} ^{a} ,
\end{equation}
where $\chi _{\xi} ^{a}= \frac{1}{2} \varepsilon ^{a} _{\hspace{1.5 mm} bc} \lambda _{\xi}^{bc} $ and $\lambda _{\xi}^{bc}$ is generator of the Lorentz gauge transformations $SO(2, 1)$, also $ \delta ^{r} _{s} $  denotes the ordinary Kronecker delta. We assume that $\xi$ may be a function of dynamical fields. In the paper \cite{3}, we have shown that quasi-local conserved charge perturbation associated with a field-dependent vector field $\xi$ is given by \footnote{We denote variation with respect to dynamical fields by $\hat{\delta}$.}
\begin{equation}\label{3}
 \hat{\delta} Q ( \xi )  = \frac{1}{8 \pi G} \int_{\Sigma} \left( \tilde{g}_{rs} i_{\xi} a^{r} - \tilde{g} _{\omega s} \chi _{\xi} \right) \cdot \hat{\delta} a^{s},
\end{equation}
where $G$ denotes the Newtonian gravitational constant and $\Sigma$ is a spacelike codimension two surface. We can take an integration of \eqref{3} over the one-parameter path on the solution space \cite{5,6} and then we find that
\begin{equation}\label{4}
  Q ( \xi )  = \frac{1}{8 \pi G} \int_{0}^{1} ds \int_{\Sigma} \left( \tilde{g}_{rs} i_{\xi} a^{r} - \tilde{g} _{\omega s} \chi _{\xi} \right) \cdot \hat{\delta} a^{s},
\end{equation}
Also, we argued that the quasi-local conserved charge \eqref{4} is not only conserved for Killing vectors which are admitted by spacetime everywhere but also it is conserved for the asymptotically Killing vectors.
\subsection{Generalized minimal massive gravity}\label{B.2}
Generalized minimal massive gravity (GMMG) is an example of the Chern-Simons-like theories of gravity \cite{7}. In the GMMG, there are four flavours of one-form, $a^{r}= \{ e, \omega , h, f \}$, and the non-zero components of the flavour metric and the flavour tensor are
\begin{equation}\label{5}
\begin{split}
     & \tilde{g}_{e \omega}=-\sigma, \hspace{1 cm} \tilde{g}_{e h}=1, \hspace{1 cm} \tilde{g}_{\omega f}=-\frac{1}{m^{2}}, \hspace{1 cm} \tilde{g}_{\omega \omega}=\frac{1}{\mu}, \\
     & \tilde{f}_{e \omega \omega}=-\sigma, \hspace{1 cm} \tilde{f}_{e h \omega}=1, \hspace{1 cm} \tilde{f}_{f \omega \omega}=-\frac{1}{m^{2}}, \hspace{1 cm} \tilde{f}_{\omega \omega \omega}=\frac{1}{\mu},\\
     & \tilde{f} _{eff}= -\frac{1}{m^{2}}, \hspace{1 cm} \tilde{f}_{eee}=\Lambda_{0},\hspace{1 cm} \tilde{f}_{ehh}= \alpha .
\end{split}
\end{equation}
where $\sigma$, $\Lambda _{0}$, $\mu$, $m$ and $\alpha$ are a sign, cosmological parameter with dimension of mass squared, mass parameter of the Lorentz Chern-Simons term, mass parameter of New Massive Gravity term and a dimensionless parameter, respectively.\\
For all the solutions of the Einstein gravity with negative cosmological constant, we have
\begin{equation}\label{6}
  R(\Omega)+\frac{1}{2l^{2}}e \times e =0 , \hspace{1 cm} T(\Omega)=0,
\end{equation}
where $R(\Omega) = d \Omega + \frac{1}{2} \Omega \times \Omega$ is curvature 2-form, $T(\Omega)= D(\Omega)e$ is torsion 2-form and $\Omega$ is torsion free spin-connection. Also, $D(\Omega)$ denotes exterior covariant derivative with respect to $\Omega$ and $l$ is AdS$_{3}$ radii. All the solutions of the Einstein gravity with negative cosmological constant solve the GMMG equations of motion when the following equations are satisfied \cite{3,8}
\begin{equation}\label{7}
  f^{a}=Fe^{a}, \hspace{1 cm} h^{a}=He^{a},
\end{equation}
\begin{equation}\label{8}
  \frac{\sigma}{ l^{2}} - \alpha (1 + \sigma \alpha ) H ^{2} + \Lambda _{0} - \frac{F^{2}}{ m^{2}}=0,
\end{equation}
\begin{equation}\label{9}
  - \frac{1}{\mu l^{2}} + 2 (1 + \sigma \alpha ) H + \frac{2 \alpha}{m^{2}} F H + \frac{\alpha ^{2}}{\mu} H^{2}=0,
\end{equation}
\begin{equation}\label{10}
  - F + \mu (1 + \sigma \alpha ) H + \frac{\mu \alpha}{m^{2}} FH=0,
\end{equation}
with
\begin{equation}\label{11}
  \omega = \Omega + \alpha h,
\end{equation}
where $F$ and $H$ are constant parameters.
\subsection{Conserved charges in the context of GMMG}\label{B.3}
One can simplify the quasi-local charge perturbation \eqref{3} in the context of GMMG for all the considered class of solutions which obey equations \eqref{6}-\eqref{11}. After some calculations, one finds that \cite{3}
\begin{equation}\label{12}
\begin{split}
   \hat{\delta} Q(\xi) = \frac{1}{8 \pi G} \int _{\Sigma} \biggl\{ & - \left( \sigma + \frac{\alpha H}{\mu} + \frac{F}{m^{2}} \right) \left( (i_{\xi} \Omega - \chi _{\xi} ) \cdot \hat{\delta} e + i_{\xi} e \cdot \hat{\delta} \Omega \right) \\
     &  + \frac{1}{\mu} \left( (i_{\xi} \Omega - \chi _{\xi} ) \cdot \hat{\delta} \Omega + \frac{1}{l^{2}} i_{\xi} e \cdot \hat{\delta} e \right) \biggr\}.
\end{split}
\end{equation}
By demanding that the Lie-Lorentz derivative of $e^{a}$ becomes zero explicitly when $\xi$ is a Killing vector field, we find the following expression for $\chi_{\xi}$ \cite{9,10}
\begin{equation}\label{13}
  \chi _{\xi} ^{a} = i_{\xi} \omega ^{a} + \frac{1}{2} \varepsilon ^{a}_{\hspace{1.5 mm} bc} e^{\nu b} (i_{\xi} T^{c})_{\nu} + \frac{1}{2} \varepsilon ^{a}_{\hspace{1.5 mm} bc} e^{b \mu} e^{c \nu} \overset{\bullet}{\nabla} _{\mu} \xi _{\nu} ,
\end{equation}
where $\overset{\bullet}{\nabla}$ denotes covariant derivative with respect to the Levi-Civita connection, and one can show that this expression can be rewritten as \cite{11}
\begin{equation}\label{14}
  i_{\xi} \Omega - \chi _{\xi} = - \frac{1}{2} \varepsilon ^{a}_{\hspace{1.5 mm} bc} e^{b \mu} e^{c \nu} \overset{\bullet}{\nabla} _{\mu} \xi _{\nu} .
\end{equation}
Alos, we mention that, the torsion free spin-connection is given by
\begin{equation}\label{15}
  \Omega ^{a} _{ \hspace{1.5 mm}\mu} = \frac{1}{2} \varepsilon^{a b c} e _{b} ^{ \hspace{1.5 mm} \alpha} \overset{\bullet}{\nabla} _{\mu} e_{c \alpha}.
\end{equation}
Thus, by using Eq.\eqref{12}, we can find the conserved charge of a given dreibein $e^{a}$ (which describes a spacetime) corresponds to a given symmetry generator $\xi$.
\section{Ba$\tilde{\text{n}}$ados geometries and their symmetries}\label{C}
The Ba$\tilde{\text{n}}$ados geometries can be expressed by the following line-element\cite{12}
\begin{equation}\label{16}
  ds^{2}=l^{2} \frac{dr^{2}}{r^{2}}-\left( r dx^{+} - \frac{l^{2} \mathcal{L}_{-}}{r} dx^{-} \right)\left( r dx^{-} - \frac{l^{2} \mathcal{L}_{+}}{r} dx^{+} \right),
\end{equation}
where $x^{\pm} = t/l \pm \phi$. $r$, $ t$ and $ \phi \sim \phi + 2 \pi$ are respectively radial, time and angular coordinates. Also, $ \mathcal{L}_{\pm} = \mathcal{L}_{\pm}(x^{\pm})$ are two arbitrary periodic functions. The line-element \eqref{16} solves equations of motion of the Einstein gravity with negative cosmological constant.\\
The metric under transformations generated by $\xi$ transforms as $\hat{\delta} _{_{\xi}} g_{\mu\nu} = \pounds _{\xi} g_{\mu\nu}$ \footnote{$\pounds _{\xi}$ denotes ordinary Lie derivative along the vector field $\xi$.}. The variation generated by the following Killing vector field preserves the form of the considered metric \cite{21}
\begin{equation}\label{17}
  \begin{split}
      \xi ^{r} = & - \frac{r}{2} \left( \partial_{+} T^{+} + \partial_{-} T^{-} \right), \\
      \xi ^{\pm} = & T^{\pm} + \frac{l^{2} r^{2} \partial _{\mp}^{2} T^{\mp}+l^{4} \mathcal{L}_{\mp} \partial _{\pm}^{2} T^{\pm}}{2 \left( r^{4} - l^{4} \mathcal{L}_{+} \mathcal{L}_{-} \right)},
  \end{split}
\end{equation}
where $ T^{\pm} =T^{\pm} (x^{\pm}) $ are two arbitrary periodic functions. In other words, under transformation generated by the Killing vector field \eqref{17}, the metric \eqref{16} transforms as \cite{21}
\begin{equation}\label{18}
  g_{\mu \nu}[\mathcal{L}_{+},\mathcal{L}_{-}] \rightarrow g_{\mu \nu}[\mathcal{L}_{+} + \hat{\delta} _{\xi} \mathcal{L}_{+} ,\mathcal{L}_{-}+ \hat{\delta} _{\xi} \mathcal{L}_{-}],
\end{equation}
with
\begin{equation}\label{19}
  \hat{\delta} _{\xi} \mathcal{L}_{\pm} = 2 \mathcal{L}_{\pm} \partial _{\pm} T^{\pm} + T^{\pm} \partial _{\pm} \mathcal{L}_{\pm} - \frac{1}{2} \partial _{\pm} ^{3} T^{\pm},
\end{equation}
where $\mathcal{L}_{\pm}$  are dynamical fields appeared in the metric \eqref{16}. As we know, Ba$\tilde{\text{n}}$ados geometries obey the standard Brown-Henneaux boundary conditions at spatial infinity \cite{18}. So, by expanding Killing vector field \eqref{17} at spatial infinity we will find asymptotic Killing vector which is presented in \cite{18}, as we expect. If we set $\hat{\delta} _{\xi} \mathcal{L}_{\pm}=0$, we will get to exact symmetries of Ba$\tilde{\text{n}}$ados geometries. In this case $T^{\pm}$ are not arbitrary functions and it has been considered in \cite{13,14}. In this paper we consider a general case in which $\hat{\delta} _{\xi} \mathcal{L}_{\pm} \neq 0$, in general.\\
Because $\xi$ depends on dynamical fields so we need to introduce a modified version of the Lie brackets \cite{15}
\begin{equation}\label{20}
  \left[ \xi_{1} , \xi_{2} \right] =  \pounds _{\xi_{1}} \xi_{2} - \delta ^{(g)}_{\xi _{1}} \xi_{2} + \delta ^{(g)}_{\xi _{2}} \xi_{1},
\end{equation}
where $\xi_{1}= \xi(T^{+}_{1}, T^{-}_{1})$ and $\xi_{2}= \xi(T^{+}_{2}, T^{-}_{2})$. In the equation \eqref{21}, $\delta ^{(g)}_{\xi _{1}} \xi_{2}$ denotes the change induced in $\xi_{2}$ due to the variation of metric $\hat{\delta} _{_{\xi _{1}}} g_{\mu\nu} = \pounds _{\xi_{\xi}} g_{\mu\nu}$ \cite{15}.
By substituting Eq.\eqref{17} into Eq.\eqref{20}, we find that
\begin{equation}\label{21}
  \left[ \xi(T^{+}_{1}, T^{-}_{1}) , \xi(T^{+}_{2}, T^{-}_{2}) \right] = \xi(T^{+}_{12}, T^{-}_{12}),
\end{equation}
where
\begin{equation}\label{22}
  T^{\pm}_{12}= T^{\pm}_{1} \partial_{\pm} T^{\pm}_{2} - T^{\pm}_{2} \partial_{\pm} T^{\pm}_{1}.
\end{equation}
Thus the algebra of the Killing vector fields is closed.
\section{"Asymptotic" and "Near Horizon" conserved charges and their algebras}\label{D}
\subsection{Asymptotic conserved charges}\label{D.1}
In this subsection we are going to obtain the conserved charges correspond to the asymptotic symmetries of generic black holes in the class of Ba$\tilde{\text{n}}$ados geometries in the context of GMMG model. Then we will obtain the algebra satisfy by these conserved charges.\\
Dreibein corresponds to the line-element \eqref{16} is given as
\begin{equation}\label{23}
  \begin{split}
     e^{0} = & \frac{1}{2} \left( r - \frac{l^{2} \mathcal{L}_{+}}{r} \right) dx^{+} + \frac{1}{2} \left( r - \frac{l^{2} \mathcal{L}_{-}}{r} \right) dx^{-}, \\
     e^{1} = & \frac{1}{2} \left( r + \frac{l^{2} \mathcal{L}_{+}}{r} \right) dx^{+} - \frac{1}{2} \left( r + \frac{l^{2} \mathcal{L}_{-}}{r} \right) dx^{-}, \\
     e^{2} =&  \frac{l}{r} dr.
  \end{split}
\end{equation}
Torsion-free spin-connection corresponds to dreibein \eqref{23} is given by
\begin{equation}\label{24}
  \begin{split}
     \Omega ^{0} = & \frac{1}{2l} \left( r - \frac{l^{2} \mathcal{L}_{+}}{r} \right) dx^{+} - \frac{1}{2l} \left( r - \frac{l^{2} \mathcal{L}_{-}}{r} \right) dx^{-} , \\
     \Omega ^{1} = & \frac{1}{2l} \left( r + \frac{l^{2} \mathcal{L}_{+}}{r} \right) dx^{+} + \frac{1}{2l} \left( r + \frac{l^{2} \mathcal{L}_{-}}{r} \right) dx^{-} , \\
     \Omega ^{2} = & 0,
  \end{split}
\end{equation}
where we used Eq.\eqref{15}. One can use Eq.\eqref{14}, Eq.\eqref{17}, Eq.\eqref{23} and Eq.\eqref{24} to find that
\begin{equation}\label{25}
  \begin{split}
      (i_{\xi} \Omega - \chi _{\xi} ) \cdot \hat{\delta} e + i_{\xi} e \cdot \hat{\delta} \Omega = & l \left( T^{+} \hat{\delta} \mathcal{L}_{+} dx^{+} - T^{-} \hat{\delta} \mathcal{L}_{-} dx^{-} \right), \\
       (i_{\xi} \Omega - \chi _{\xi} ) \cdot \hat{\delta} \Omega + \frac{1}{l^{2}} i_{\xi} e \cdot \hat{\delta} e = & T^{+} \hat{\delta} \mathcal{L}_{+} dx^{+} + T^{-} \hat{\delta} \mathcal{L}_{-} dx^{-}.
  \end{split}
\end{equation}
By substituting Eq.\eqref{25} into Eq.\eqref{12} and taking an integration over one-parameter path on the solution space, we obtain conserved charge corresponds to the Killing vector field \eqref{17} as
\begin{equation}\label{26}
  Q(\xi)=Q^{+}(T^{+})+Q^{-}(T^{-}),
\end{equation}
where
\begin{equation}\label{27}
  Q^{\pm}(T^{\pm})=\mp \frac{l}{8 \pi G}\left( \sigma + \frac{\alpha H}{\mu} + \frac{F}{m^{2}} \mp \frac{1}{\mu l} \right) \int_{\Sigma} T^{\pm} \mathcal{L}_{\pm} dx^{\pm}.
\end{equation}
It is clear now that the spacelike codimension two surface $\Sigma$ can be taken a circle of arbitrary radius, and this is a consequence of quasi-local formalism.\\
The algebra of conserved charges can be written as \cite{16}
\begin{equation}\label{28}
  \left\{ Q(\xi _{1}) , Q(\xi _{2}) \right\} = Q \left(  \left[ \xi _{1} , \xi _{2} \right] \right) + \mathcal{C} \left( \xi _{1} , \xi _{2} \right)
\end{equation}
where $\mathcal{C} \left( \xi _{1} , \xi _{2} \right)$ is central extension term. Also, the left hand side of the equation \eqref{28} can be defined by
\begin{equation}\label{29}
  \left\{ Q(\xi _{1}) , Q(\xi _{2}) \right\}= \hat{\delta} _{\xi _{2}} Q(\xi _{1}).
\end{equation}
Therefore one can find the central extension term by using the following formula
\begin{equation}\label{30}
  \mathcal{C} \left( \xi _{1} , \xi _{2} \right)= \hat{\delta} _{\xi _{2}} Q(\xi _{1}) - Q \left(  \left[ \xi _{1} , \xi _{2} \right] \right).
\end{equation}
By substituting Eq.\eqref{19}, Eq.\eqref{21} and Eq.\eqref{26} into Eq.\eqref{30} we obtain the central extension term, then by substituting obtained result into Eq.\eqref{28} we have
\begin{equation}\label{31}
\begin{split}
   \left\{ Q^{\pm}(T^{\pm}_{1}) , Q^{\pm}(T^{\pm}_{2}) \right\} = & Q^{\pm}(T^{\pm}_{12}) \\
    + \frac{l}{4 \pi G}\left( \sigma + \frac{\alpha H}{\mu} + \frac{F}{m^{2}} \mp \frac{1}{\mu l} \right) \biggl\{ &  \int_{\Sigma} T^{\pm}_{12} \mathcal{L}_{\pm} dx^{\pm}\\
     & -\frac{1}{8} \int_{\Sigma} \left( T^{\pm}_{1} \partial _{\pm}^{3} T^{\pm}_{2} - T^{\pm}_{2} \partial _{\pm}^{3} T^{\pm}_{1} \right) dx^{\pm} \biggr\},
\end{split}
\end{equation}
\begin{equation}\label{32}
  \left\{ Q^{\pm}(T^{\pm}_{1}) , Q^{\mp}(T^{\mp}_{2}) \right\} =0.
\end{equation}
By introducing Fourier modes $ Q^{\pm} _{m} = Q^{\pm} (e^{imx^{\pm}})$, we find that
\begin{equation}\label{33}
\begin{split}
   i \left\{ Q^{\pm} _{m} , Q^{\pm} _{n} \right\} =  & (m-n) Q^{\pm} _{m+n} \\
     +\frac{l}{8 G} & \left( \sigma + \frac{\alpha H}{\mu} + \frac{F}{m^{2}} \mp \frac{1}{\mu l} \right) \left\{ n^{3} \delta_{m+n,0} + \frac{2}{\pi}
     (m-n) \tilde{\mathcal{L}}_{\pm(m+n)} \right\},
\end{split}
\end{equation}
\begin{equation}\label{34}
   i \left\{ Q^{\pm} _{m} , Q^{\mp} _{n} \right\} =0,
\end{equation}
where
\begin{equation}\label{35}
  \tilde{\mathcal{L}}_{\pm m} = \int_{\Sigma} e^{i m x^{\pm}} \mathcal{L}_{\pm} dx^{\pm}.
\end{equation}
Now we set $ \hat{L}^{\pm} _{m} \equiv Q^{\pm} _{m}$ and replace the Dirac brackets by commutators $i \{ , \} \rightarrow [,]$. After making a constant shift on the spectrum of $\hat{L}^{\pm} _{m} $ \cite{17}, we find that
\begin{equation}\label{36}
  \left[ \hat{L}^{\pm} _{m},\hat{L}^{\pm} _{n} \right]=(m-n)\hat{L}^{\pm} _{m+n}+ \frac{c_{\pm}}{12} n^{2}(n-1) \delta_{m+n,0},
\end{equation}
\begin{equation}\label{37}
  \left[ \hat{L}^{+} _{m},\hat{L}^{-} _{n} \right]=0,
\end{equation}
where $c_{\pm}$ are central charges and they are given by
\begin{equation}\label{38}
  c_{\pm}= \frac{3l}{2 G} \left( \sigma + \frac{\alpha H}{\mu} + \frac{F}{m^{2}} \mp \frac{1}{\mu l} \right).
\end{equation}
It is clear that $\hat{L}^{\pm} _{m}$ are generators of the Virasoro algebra and then the algebra among the asymptotic conserved charges is isomorphic to two copies of the Virasoro algebra.
\subsection{Near horizon conserved charges}\label{D.2}
One can use the method presented in Sec.\ref{B} to find conserved charges of near horizon geometries like what presented in \cite{19}, which obey the near horizon fall-off conditions of non-extremal black holes in three dimension. We did this work in the paper \cite{3} and we found the near horizon conserved charges in the context of GMMG. Also, we showed that the obtained near horizon conserved charges obey the following algebra
\begin{equation}\label{39}
  \left[ \hat{J}^{\pm} _{m},\hat{J}^{\pm} _{n} \right]= \mp \frac{c_{\pm}}{12} m \delta_{m+n,0} \hspace{2 mm},
\end{equation}
\begin{equation}\label{40}
  \left[ \hat{J}^{+} _{m},\hat{J}^{-} _{n} \right]= 0.
\end{equation}
Similar to the near horizon symmetry algebra in the Einstein gravity with negative cosmological constant \cite{19}, the algebra \eqref{39} and \eqref{40} consists of two $U(1)$ current algebras, but instead with levels $\pm \frac{l}{8 G}$, here the level of algebra is given by $\mp \frac{c_{\pm}}{12}$.
\subsection{Relation Between Asymptotic and Near Horizon Algebras}\label{D.3}
To relate the asymptotic Virasoro algebra \eqref{36} and the near horizon algebra \eqref{39}, we need a twisted Sugawara construction \cite{20} as follows:
\begin{equation}\label{41}
  \hat{\mathfrak{L}}^{\pm}_{m} = \frac{i m}{\sqrt{\pm 2}} \hat{J}^{\pm} _{m} \mp \frac{6}{c_{\pm}} \sum_{p \in \mathbb{Z}} \hat{J}^{\pm} _{m-p} \hat{J}^{\pm} _{p}.
\end{equation}
It is straightforward to show that
\begin{equation}\label{42}
  \left[ \hat{\mathfrak{L}}^{\pm} _{m},\hat{\mathfrak{L}}^{\pm} _{n} \right]=(m-n)\hat{\mathfrak{L}}^{\pm} _{m+n}+ \frac{c_{\pm}}{12} n^{3} \delta_{m+n,0},
\end{equation}
Now, we can get to the asymptotic Virasoro algebra by making a shift on the spectrum of $\hat{\mathfrak{L}}^{\pm} _{m}$ by a constant,
\begin{equation}\label{43}
  \hat{L}^{\pm} _{m} = \hat{\mathfrak{L}}^{\pm}_{m}+ \frac{c_{\pm}}{24} \delta_{m+n,0} \hspace{2 mm}.
\end{equation}
In this way, we could relate near horizon symmetry algebra to the asymptotic one in the context of GMMG model.
\section{Conclusion}\label{E}
 In the Chern-Simons-like theories of gravity, the off-shell quasi-local conserved charge corresponds to (asymptotic) Killing vector $\xi$ is given by Eq.\eqref{4}. As an example of CSLTs, in this paper we have considered GMMG model. All the solutions of the Einstein gravity with negative cosmological constant are solutions of GMMG (see subsection \ref{B.2}). One can simplify the off-shell quasi-local conserved charge corresponds to ( the asymptotic) Killing vector $\xi$ as Eq.\eqref{12} for considered solutions in the context of GMMG model. In sec.\ref{C}, we have considered the Ba$\tilde{\text{n}}$ados geometries which are given by the line-element \eqref{16}. We saw that the Killing vector field \eqref{17} preserves the form of metric \eqref{16}, i.e. under transformation generated by the Killing vector field \eqref{17}, the metric \eqref{16} transforms as $g_{\mu \nu}(\Phi) \rightarrow g_{\mu \nu}(\Phi + \hat{\delta}_{\xi} \Phi)$, where $\Phi=\{ \mathcal{L}_{+},\mathcal{L}_{-} \}$. The algebra of the Killing vectors \eqref{17} is closed with respect to a modified version of the Lie brackets (see Eq.\eqref{20} and Eq.\eqref{21}). In subsection \ref{D.1}, we have found the asymptotic conserved charges of the Ba$\tilde{\text{n}}$ados geometries using the off-shell quasi-local approach. As we have seen, it does not matter that the integration surface is located at spatial infinity or it to be a circle of finite radius. Also, in subsection \ref{D.1}, we have shown that the algebra among the asymptotic conserved charges is isomorphic to two copies of the Virasoro algebra (see Eq.\eqref{36} and Eq.\eqref{37}) with central charges $c_{\pm}$. As we have mentioned in subsection \ref{D.2}, the algebra of near horizon conserved charges is given by Eq.\eqref{39} and Eq.\eqref{40}. Eventually, in subsection \ref{D.3}, we have related the algebra of near horizon conserved charges to the asymptotic one. So, in this paper we have shown that the main part of the horizon fluffs proposal of Refs.\cite{140,14} appear for generic black holes in the class of Ba$\tilde{\text{n}}$ados geometries in the context of GMMG model. In the other words, we have shown that asymptotic conserved charges satisfy two copies of the Virasoro algebra in one hand, and in another hand by using a twisted Sugawara construction, we have related the near horizon symmetry algebra to the asymptotic one.

\end{document}